\documentclass{iaus}
\usepackage{graphicx}

\title[Dynamics, Origin, and Activation of MBCs] 
{Dynamics, Origin, and Activation of \\ Main Belt Comets}

\author[N. Haghighipour]  
{Nader Haghighipour$^1$}

\affiliation{$^1$Institute for Astronomy and NASA Astrobiology Institute, University of Hawaii, 2680 Woodlawn Drive, Honolulu, HI 96822, USA \\email: {\tt nader@ifa.hawaii.edu}}

\pubyear{2009}
\volume{263}  
\pagerange{1-6}
\jname{Icy Bodies of the Solar System}
\editors{J. Fernandez et al., eds.}
\begin{document}

\maketitle

\begin{abstract}
The discovery of Main Belt Comets (MBCs) has raised many questions regarding the origin
and activation mechanism of these objects. Results of a study of the  dynamics of these 
bodies suggest that MBCs were formed in-situ as the remnants of the break-up of large
icy asteroids. Simulations show that similar to the asteroids in the main belt, MBCs 
with orbital eccentricities smaller than 0.2 and inclinations lower than $25^\circ$ have 
stable orbits implying that many MBCs with initially larger eccentricities and inclinations
might have been scattered to other regions of the asteroid belt. Among scattered MBCs, 
approximately 20\% reach the region of terrestrial planets where they might have
contributed to the accumulation of water on Earth. Simulations also show that collisions
among MBCs and small objects could have played an important role in triggering the
cometary activity of these bodies. Such collisions might have exposed sub-surface water ice
which sublimated and created thin atmospheres and tails around MBCs.
This paper discusses the results of numerical studies of the dynamics
of MBCs and their implications for the origin of these objects. 
The results of a large numerical modeling of the collisions of m-sized bodies 
with km-sized asteroids in the outer part of the asteroid belt are also presented and the viability 
of the collision-triggering activation scenario is discussed. 
\end{abstract}

\firstsection 
\section{Introduction}

The discovery of comet-like activities in four icy asteroids 7968 Elst-Pizzaro (133P/Elst-Pizzaro), 
118401 (1999 ${{\rm RE}_{70}}$, 176P/LINEAR), P/2005 U1 (Read), and P/2008 R1 (Garradd) 
has added a new item to the mysteries of the asteroid belt (\cite[Hsieh \& Jewitt 2006]{Hsieh06};
\cite[Jewitt, Yang \& Haghighipour 2009]{Jewitt09}). Known as Main Belt Comets (MBCs), 
these objects may be representatives of a new class of bodies
that are dynamically asteroidal (i.e., their Tisserand parameters\footnote{For a small 
object, such as an asteroid, that is subject to the gravitational 
attraction of a central star and the perturbation of a planetary 
body P, the quantity ${a_{\rm P}}/a \, + \,2 {[(1-e^2)\,a/{a_{\rm P}}]^{1/2}}\, \cos i$ 
is defined as its Tisserand parameter . In this formula, $a$ is the
semimajor axis of the object with respect to the star, $e$ is its 
orbital eccentricity, $i$ is its orbital inclination, and $a_{\rm P}$ 
is the semimajor axis of the planet. In the solar system, the 
Tisserand parameter of a small body with respect to Jupiter can be used 
to determine the cometary or asteroidal nature of its orbit. In general, 
the Tisserand parameters of comets with respect to Jupiter are smaller 
than 3, whereas those of asteroids are mostly larger.}  are larger than 3), 
but have cometary appearance. As shown in Table 1, the orbits of these 
objects are in the outer half of the asteroid belt (Fig. 1) implying 
that they may contain sub-surface water ice. In fact the observation of the
tail of 7968 Elst-Pizzaro by \cite[Hsieh, Jewitt \& Fern\'andez (2004)]{Hsieh04}
has indicated that the comet-like activity of this MBC is episodic (it is not the 
ejection of dust particles that were produced through an impact to this object) and is due to 
the dust particles that have been blown off the surface of this body by the 
drag force of the gas that was most likely produced by the sublimation of near-surface 
water ice.

\begin{table}
  \begin{center}
  \caption{Orbital Elements of MBCs (\cite[Hsieh \& Jewitt 2006]{Hsieh06})}
  \label{tab1}
 {\scriptsize
  \begin{tabular}{|l|c|c|c|c|c|}\hline 
{\bf MBC} & {\bf $a$ (AU)} & {\bf $e$} & {\bf $i$ (deg.)} & {\bf Tisserand} & {\bf Diameter (km)} \\ \hline
(133P)/7968 Elst-Pizzaro     & 3.156  & 0.165  & 1.39  & 3.184  & 5.0  \\ \hline
118401 (176P/LINEAR) & 3.196  & 0.192  & 0.24  & 3.166  & 4.4 \\ \hline
P/2005 U1 (Read) & 3.165  & 0.253  & 1.27  & 3.153  & 0.6\\ \hline
P/2008 R1 (Garradd) & 2.726 & 0.342 & 15.9 & 3.216 & 1.4 \\ \hline
  \end{tabular}
  }
 \end{center}
\end{table}

The comet-like appearance of MBCs has raised questions regarding the origin 
of these objects. While the asteroidal orbits of these bodies, combined with the
proximity of 7968 Elst-Pizzaro, 118401 (176P/LINEAR), and P/2005 U1 (Read) to
the Themis and Beagle families of asteroids (Fig. 1), suggests that MBCs have formed in-situ
as the remnants of collisionally broken larger objects, the cometary activities of these 
bodies may be taken to argue that MBCs are comets that were scattered inward from
the outer regions of the solar system and were captured in their current orbits. Such a 
capture mechanism could not have occurred recently. Simulation of the dynamics of 
Kuiper belt object by \cite[Fernandez et al. (2002)]{Fernandez02} have shown that,
at the current dynamical state of the solar system, it would not be possible to scatter 
comets from regions outside the orbit of Neptune to the main asteroid belt. A primordial 
capture, on the other hand, may not be impossible. Recently 
\cite[Levison et al. (2009)]{Levison09} have shown that within the context of
the Nice model (\cite[Gomes et al. 2005]{Gomes05}; \cite[Morbidelli et al. 2005]{Morbidelli05}; 
\cite[Tsiganis et al. 2005]{Tsiganis05}) many trans-Neptunian objects could  have 
been scattered inwards and captured in orbits in the asteroid belt as close in as 2.68 AU 
during the early state of the dynamics of the solar system. Whether these objects could be the
source of MBCs is, however, uncertain. This paper evaluates this possibility, in particular in
comparison with the in-situ formation model,  by presenting the results of a numerical 
study if the dynamics of the currently known MBCs, and discussing their implications
for the formation and origin of these objects.

As mentioned above, tails of MBCs are generated through the interactions of dust grains 
on the surface of these bodies with the gas produced by the sublimation
of near-surface water ice (\cite[Hsieh, Jewitt \& Fern\'andez 2004]{Hsieh04}). 
As shown by  \cite[Schorghofer (2008)]{Schorghofer08}, 
asteroids in the region between 2 AU and 3.3 AU can maintain sub-surface
water ice for several billion years if their surfaces are covered by a layer of dust, 
even as thick as only a few meters. That implies that in order for an MBC to start its 
cometary activation, this dusty layer has to be removed. \cite[Hsieh \& Jewitt (2006)]{Hsieh06}
have suggested that collisions between MBCs and objects as small as a few meter
in size, can reveal the sub-surface water ice. Such collisions will result in the local exposure
of ice which sublimates and creates a thin atmosphere and tail for an MBC. An activated MBC,
on the other hand, may terminate its activity after sublimating all the
ice at the location of its collision with a m-sized projectile. It may also start new activation
if it is collided with a second m-sized object in a later time again. In other words, an MBC may
be activated several times till it exhausts all its water ice, or is scattered to an unstable
orbit and either leaves the asteroid belt or collides with another asteroid or a planet.
It will therefore be useful to study the rate of the collisions of m-size bodies with km-size
MBCs, in particular in the outer region of the asteroid belt. This paper presents the results
of such simulations and discusses their implications for the activation of MBCs and
the possibility of the detection of more of these objects.

\begin{figure}
%\begin{figure}[b]
\vspace*{0.2 cm}
\begin{center}
\includegraphics[width=4in]{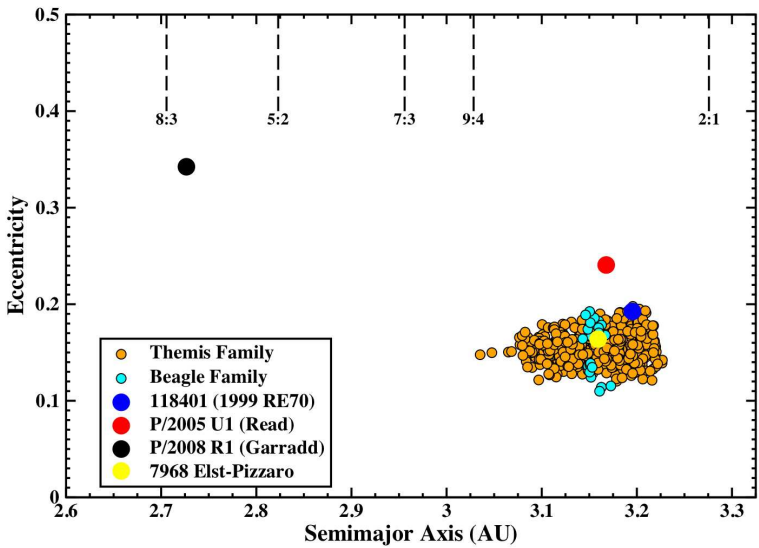} 
\caption{The four currently known MBCs and the Themis and Beagle families of 
asteroids. As shown here, 7968 Elst-Pizzaro and 118401 (Read) are within the Themis
and Beagle families whereas P/2005 U1 is in their proximity. The MBC P/2008 R1 (Garradd)
seems to be an object that was scattered out of its forming region. The locations of 
mean-motion resonances with Jupiter are also shown.}
\label{fig1}
\end{center}
\end{figure}

\section {Orbital Integrations and Implications for the Origin of MBCs}

To study the long-term stability of the four known MBCs, the orbits of these objects
were integrated for 1 Gyr. Integrations included all the planets and Pluto, and
treated MBCs as non-interacting objects. The effects of non-gravitational 
forces such as Yarkovsky, and the effect of the mass-loss of MBCs 
due to their cometary activities were not included. Since the activation of MBCs 
is episodic and intensifies during the perihelion passages of these objects, which is
short compared to their orbital periods, the effect of the mass-loss may not alter 
the dynamics of these objects significantly.
Integrations were carried out with Bulirsch-Stoer 
and with the Second-Order Mixed-Variable Symplectic (MVS)
integrators in the N-body integration package MERCURY (\cite[Chambers 1999]{Chambers99}).
The initial orbital elements of the MBCs and planets were obtained from 
documentation on solar system dynamics published by the Jet Propulsion Laboratory 
(http://ssd.jpl.nasa.gov/?bodies). The timestep of each integration was set to 9 days.

Figure 2 shows the results of the simulations. As shown here
7968 Elst-Pizzaro and 118401 (176P/LINEAR) maintain their orbits 
for 1 Gyr. However, P/2005 U1 (Read) and P/2008 R1 (Garradd) 
become unstable in approximately 20 Myr. 
Integrations were also carried out for different initial values of the semimajor axes and 
eccentricities of MBCs, changing these quantities
in increments of $\Delta a=0.0001$ AU and $\Delta e= 0.001$
within the ranges of their observational uncertainties.
Similar results were obtained. 7968 Elst-Pizzaro and 118401 were stable whereas
P/2005 U1 and P/2008 R1 became unstable in all simulations with a median lifetime of 
$\sim$57 Myr. For more details on the results of the simulations, in particular on the analysis
of the effects of mean-motion resonances on the dynamics of these MBCs,
the reader is referred to \cite[Haghighipour (2009)]{Hagh09} and
\cite[Jewitt, Yang \& Haghighipour (2009)]{Jewitt09}.

\begin{figure}
%\begin{figure}[b]
\vspace*{0.2cm}
\begin{center}
\includegraphics[width=2.5in]{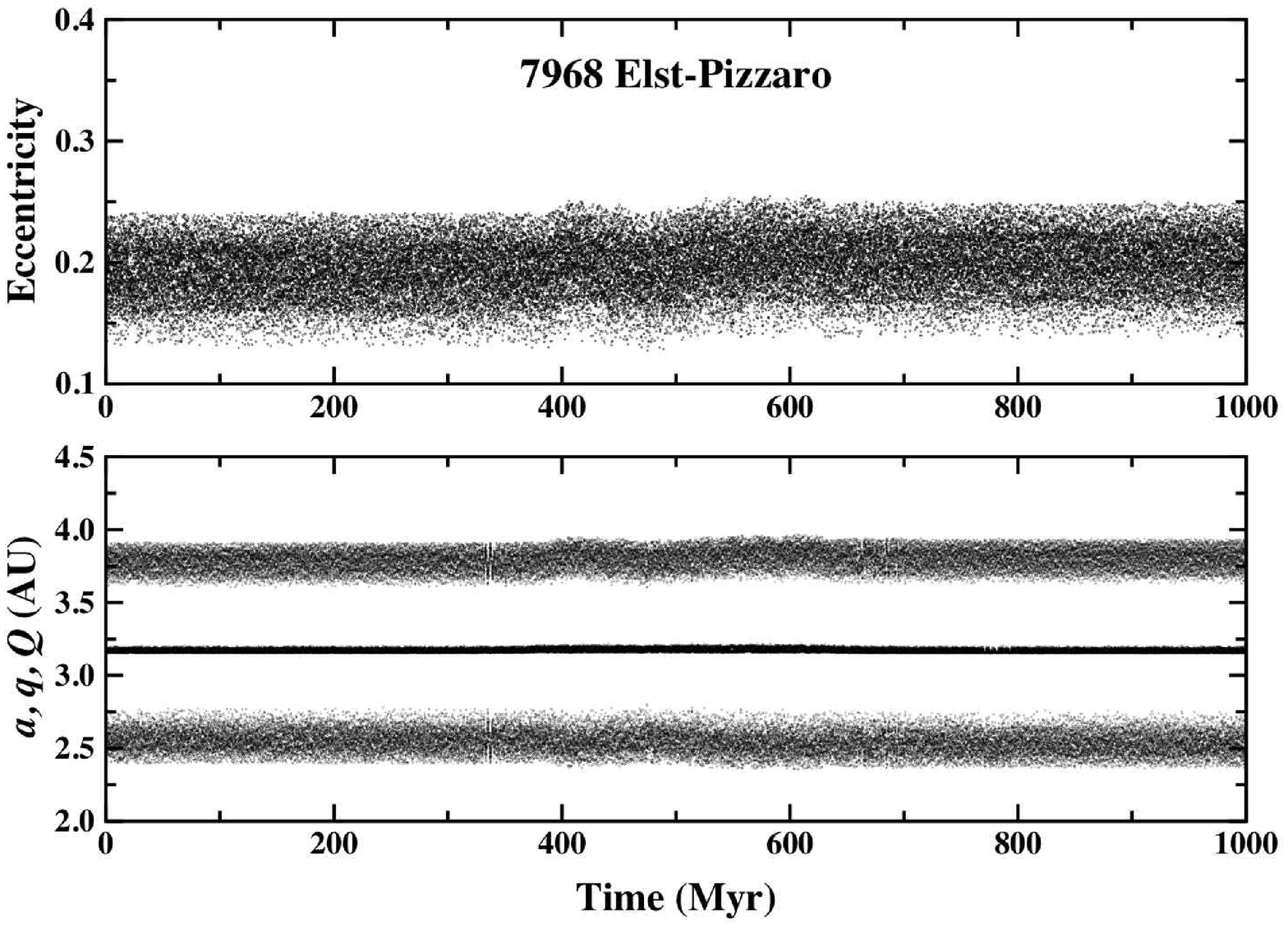}
 \includegraphics[width=2.5in]{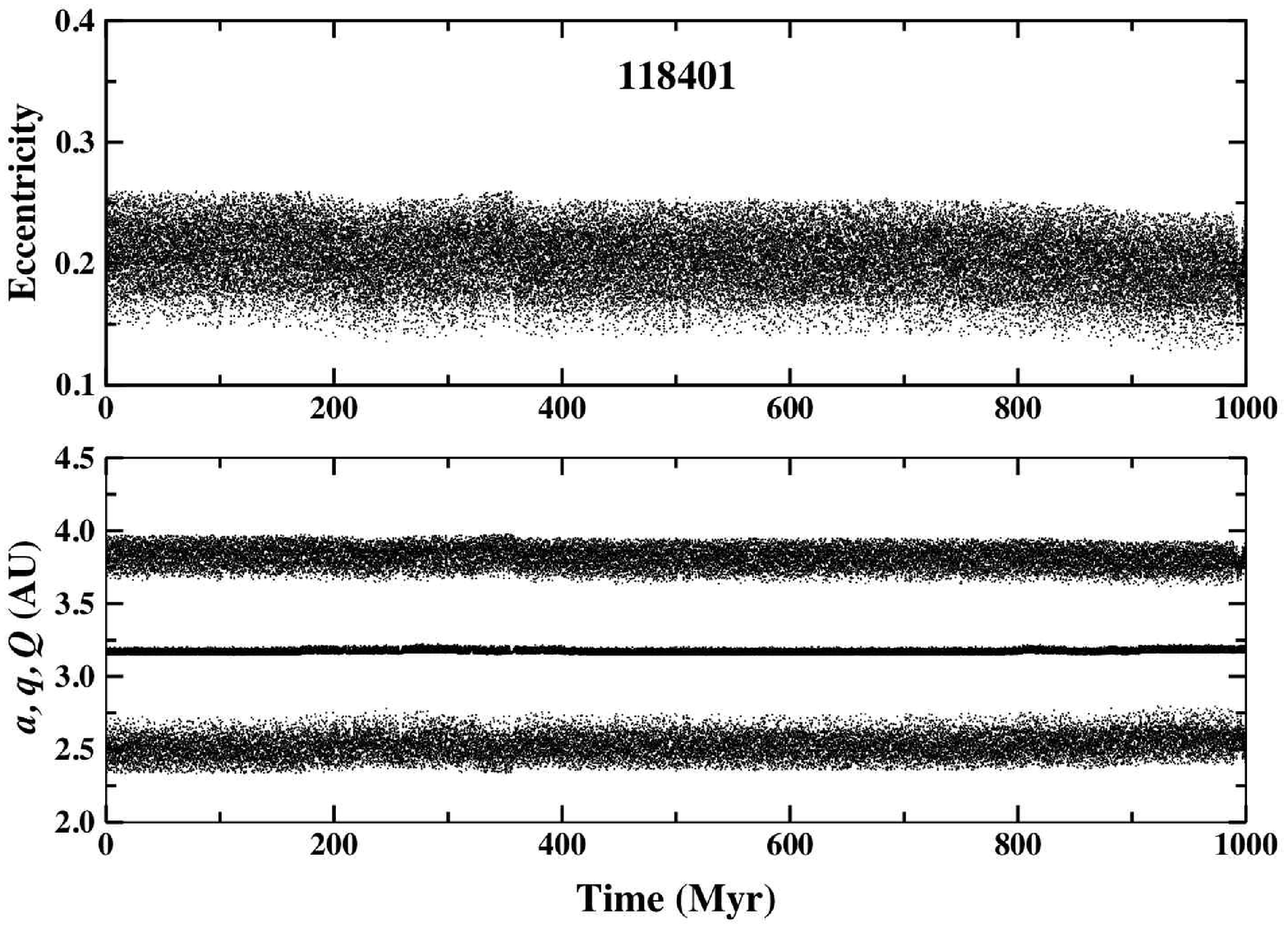}
 \includegraphics[width=2.5in]{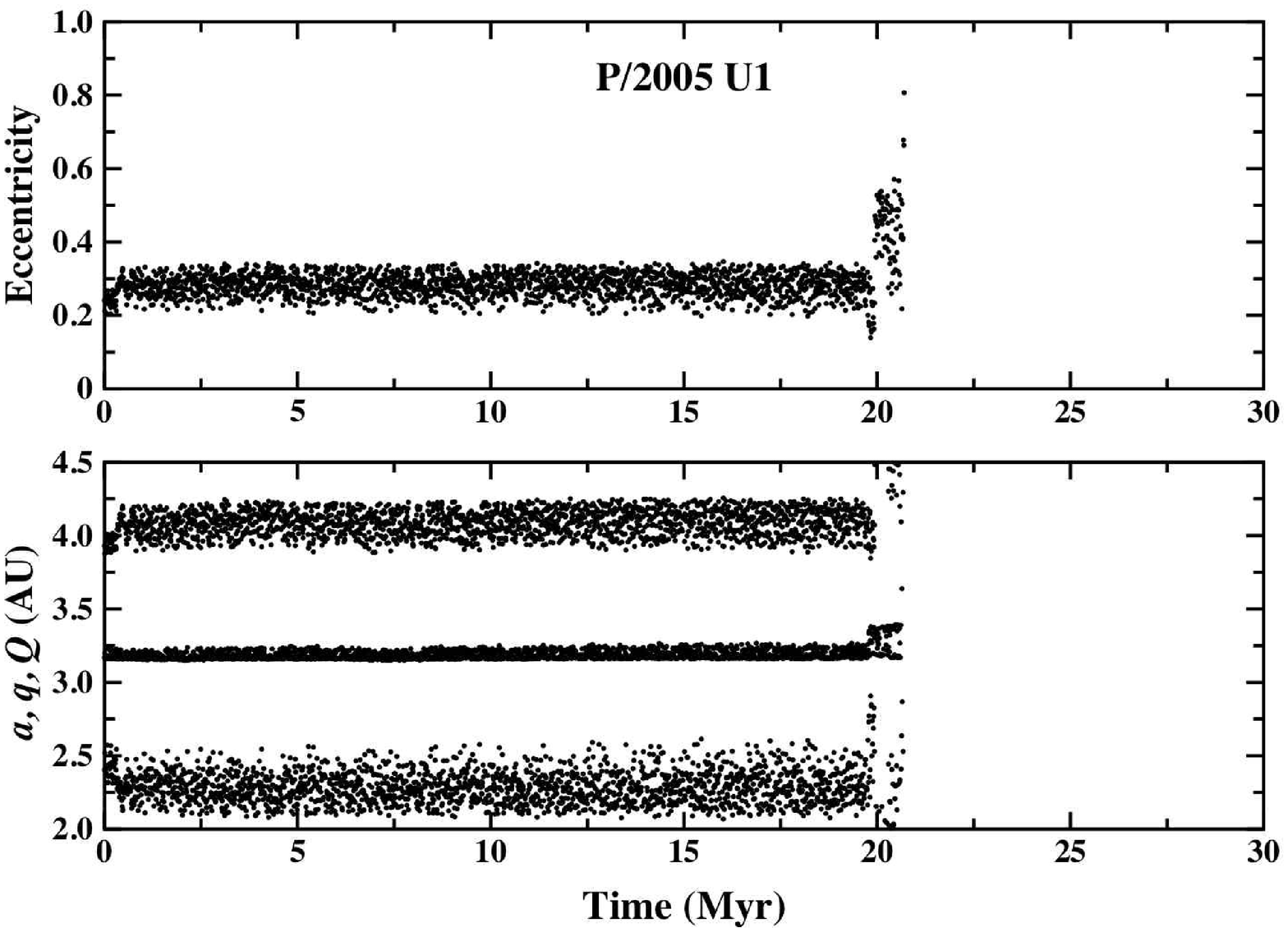}
 \includegraphics[width=2.5in]{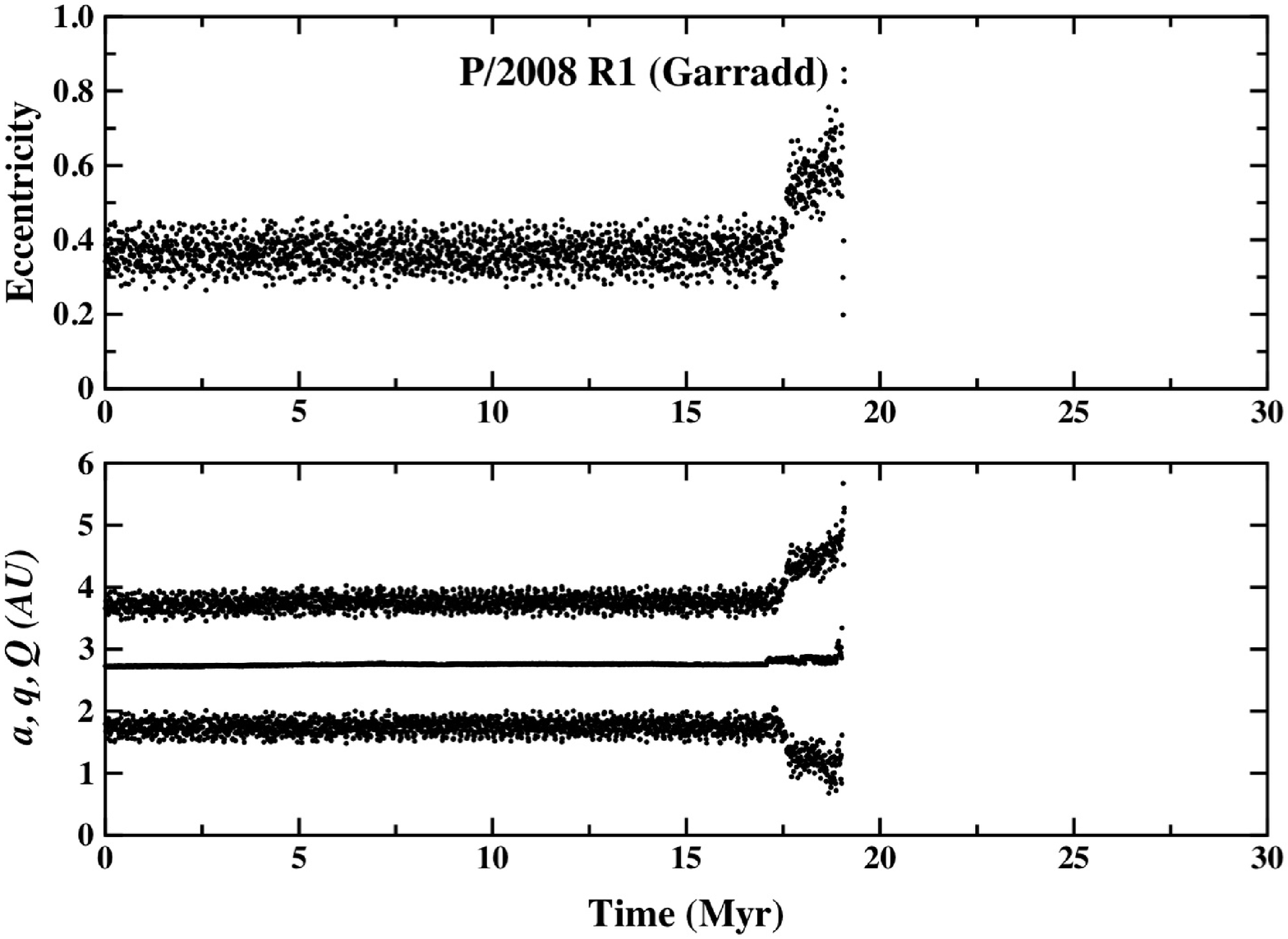}
 % \vspace*{-1.0 cm}
 \caption{Graphs of the eccentricities, semimajor axes $(a)$,
perihelion $(q)$, and aphelion $(Q)$ distances of 
7968 Elst-Pizzaro, 118401 (176P/LINEAR), P/2005 U1 (Read), 
and P/2008 R1 (Garradd) . As shown here,
7968 Elst-Pizzaro and 118401 are stable for 1 Gyr whereas P/2005 U1 (Read)
and P/2008 R1 (Garradd) become unstable in approximately 20 Myr.}
\label{fig2}
\end{center}
\end{figure}

As shown by Fig. 1, the orbit of P/2008 R1 (Garradd) is close to the influence
zone of the 8:3 mean-motion resonance with Jupiter. Numerical simulations by
\cite[Jewitt, Yang \& Haghighipour (2009)]{Jewitt09} have shown that 
the region in the vicinity of P/2008 R1 is dynamically unstable implying that this MBC must 
have formed in another region of the asteroid belt and scattered to its current orbit.
The orbital instability of P/2005 U1 (Read), on the other hand, 
may show a pathway to such scattering events.
The proximity of P/2005 U1 to the Themis family and the location of the 1:2 mean-motion
resonance with Jupiter suggest that this MBC was perhaps formed close to the
influence zone of  the 1:2 resonance. The original proximity of P/2005 U1 to 
this resonance has resulted in a gradual increase in its orbital eccentricity which will 
eventually make its orbit unstable. Such an instability might have also happened to 
the orbits of other MBCs and resulted in their scattering to other regions.
To study this scenario, a large number of hypothetical MBCs were considered around the region
where 7968 Elst-Pizzaro, 118401 (176P/LINEAR), and P/2005 U1 (Read) exist. The semimajor 
axes of these objects were varied between 3.14 AU and 3.24 AU, and their 
initial eccentricities were taken to be between 0 and 0.4. The initial orbital inclinations 
of these MBCs were chosen from a range of 0 to $40^\circ$. 

The orbits of these hypothetical MBCs were integrated for 100 Myr.
Figure 3 shows the results. In this figure, green circles correspond to MBCs 
with stable orbits whereas purple indicates instability. As shown here, 
7968 Elst-Pizzaro and 118401 (176P/LINEAR) are in the stable region of the
graph whereas P/2005 U1 (Read) is approaching the unstable area.

\begin{figure}
\vspace*{0.2 cm}
\begin{center}
 \includegraphics[width=4in]{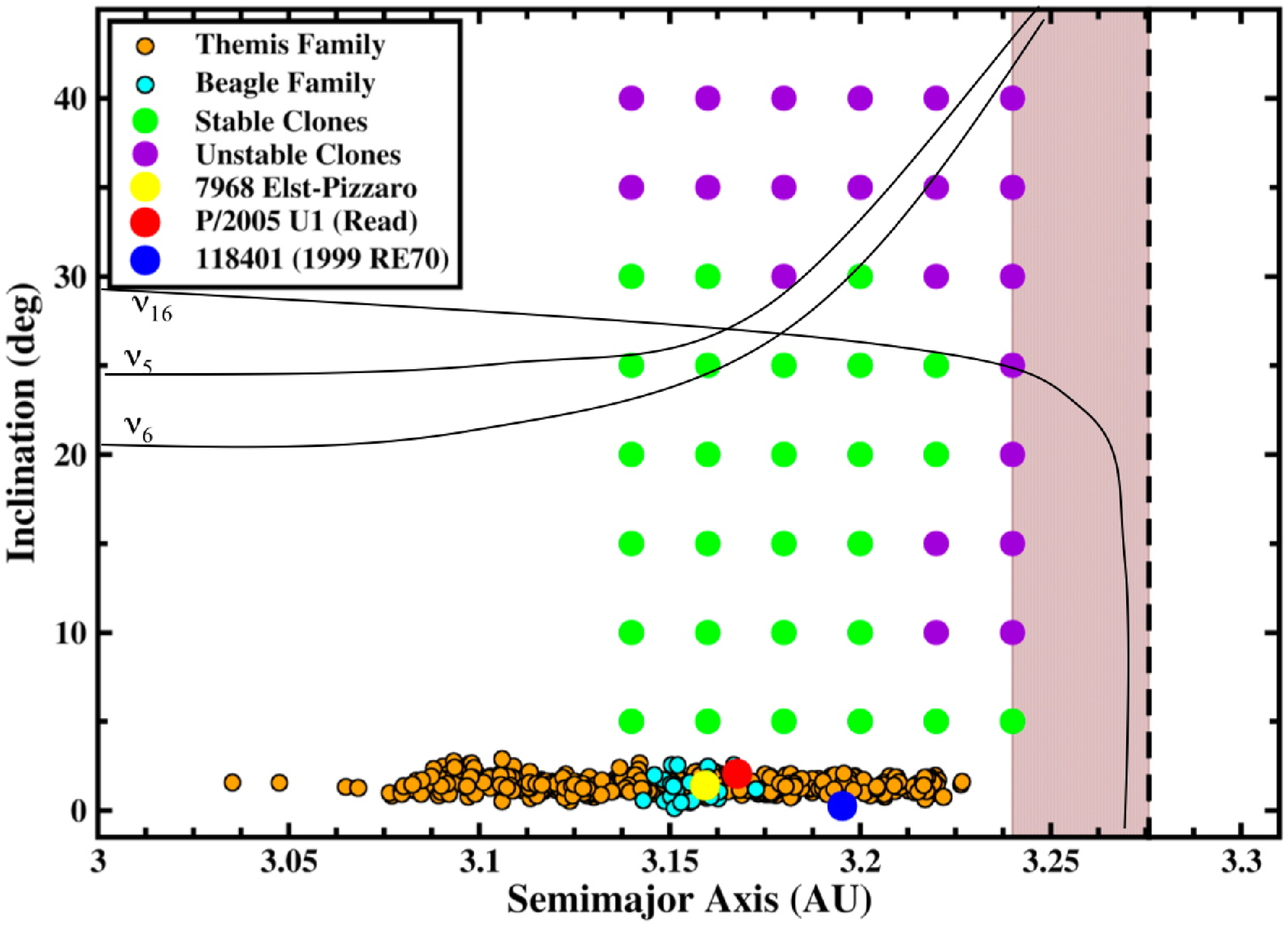} 
 \vskip 10pt
  \includegraphics[width=4in]{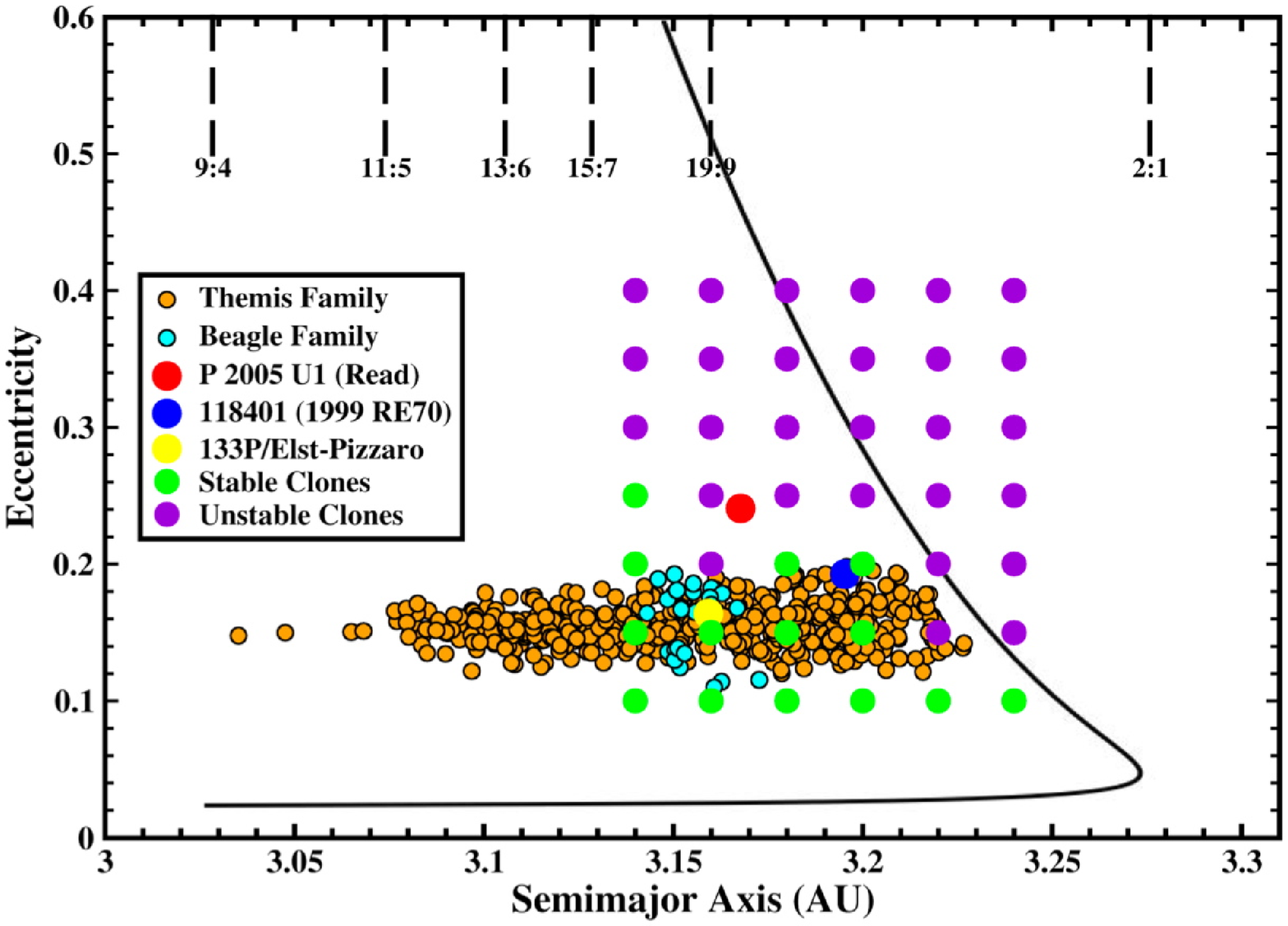} 
% \vspace*{-1.0 cm}
 \caption{Top: graph of the stability of hypothetical MBCs 
in terms of their inclinations. 
The regions of secular resonances $\nu_5$, $\nu_6$, and $\nu_{16}$
corresponding to an eccentricity of 0.1 are also shown.
Bottom: graph of the stability of hypothetical MBCs
in terms of their eccentricities. 
The brown area in the top graph and solid line in the bottom graph show 
the region of the 2:1 MMR with Jupiter. Circles in green correspond
to initial semimajor axes and eccentricities of stable MBC whereas those in purple 
show instability. Similar to the asteroid in the main belt, objects with inclinations larger 
than $\sim25^\circ$ and eccentricities larger than $\sim 0.2$ are unstable.}
   \label{fig3}
\end{center}
\end{figure}

An interesting result depicted by Fig. 3 is the familiar role of secular resonances in 
establishing the boundaries of stable zones. 
Similar to the asteroids in the asteroid belt, stability of an MBC
depends on the values of its initial eccentricity and orbital inclination. Fig.  3 shows that
for initial inclinations larger than $\sim 25^\circ$, the orbit of an MBC becomes
unstable due to the Kozai and the $\nu_5$, $\nu_6$ and $\nu_{16}$ secular resonances.
For smaller values of inclination,  the apastron distance of an MBC determines its stability.
Those hypothetical MBCs close to or inside the region of the 2:1 MMR with Jupiter became 
unstable in a short time. An analysis of the orbits of the unstable objects 
indicates that approximately $80\%$ of these bodies were scattered to 
large distances outside the solar system. This is a familiar result that 
has also been reported by \cite[O'Brien et al. (2007)]{OBrien07} and 
\cite[Haghighipour \& Scott (2008)]{Hagh08} in their
simulations of the dynamical evolution of planetesimals in the outer 
asteroid belt. From the remaining $20\%$ unstable MBCs, approximately $15\%$ 
collided with Mars, Jupiter, or Saturn, and a small fraction $(\sim 5\%)$ 
reached the region of 1 AU implying that MBCs might have played a role in delivering
water to the Earth.

The stability analysis above has direct implications for the origin of MBCs and favors
the in-situ formation of these objects. In this scenario,
MBCs are small asteroidal bodies that were formed 
as a result of the collisional break-up of their larger precursor asteroids. 
An alternative scenario based on the primordial capture of cometary bodies, 
although, as shown by \cite[Levison et al. (2009)]{Levison09}, efficient
in the inward scattering of D-type and P-type asteroids and the delivery of these objects 
in particular to the region of Trojans, cannot provide information on the inward scattering
and distribution of C-type asteroids. That is primarily due to the fact that C-type asteroids are mainly 
at small semimajor axes, and the difference between their orbital distribution
and that of D-type asteroids are not known. Additionally, the colorless feature of
MBCs, as indicated by \cite[Hsieh \& Jewitt (2006)]{Hsieh06} and 
\cite[Hsieh, Jewitt \& Fern\'andez (2008)]{Hsieh08} is not consistent with an origin
model based on the inward scattering of comets from the Kuiper belt region (the
latter objects are optically red). 

The in-situ formation scenario
is, however, consistent with MBCs orbital and spectral properties.
In this scenario, the break up of the precursor asteroids could have produced 
many km-sized fragments, among which those with large
inclinations and large eccentricities became unstable and were scattered to
other regions. The remaining objects have naturally asteroidal  orbits
(i.e. their Tisserand numbers are larger than 3), and similar to their parent bodies,
are C-type asteroids with no specific optical color. In regard to  7968 Elst-Pizzaro, 
118401 (176P/LINEAR), and P/2005 U1 (Read), this scenario points to the 
Themis family, and perhaps a smaller $\sim 10$ Myr sub-family (known as Beagle) 
within these objects (\cite[Nesvorn\'y et al 2008]{Nesvorny08}), as the origin of these MBCs. 
This scenario also suggest that asteroid families, in particular those in the outer
half of the asteroid belt and with large parent bodies capable of differentiating and 
forming ice-rich mantles, are the most probable places for detecting more MBCs.
As indicated by the results of the dynamical simulations, some members of such 
families may interact with giant planets and reach 
orbits in other regions of the asteroid belt--a scenario that might explain the 
existence of P/2008 R1 (Garradd) in its current orbit. All sky surveys such as those with
Pan STARRS 1 would be capable of detecting such individual MBCs, and are ideal for 
carrying out targeted surveys for families of these objects.

\section{Collision With Small Objects As The Activation-Triggering Mechanism}

As mentioned in the introduction, it has been suggested that the tails of MBCs are 
dust particles that have been carried away from the surfaces of these object by the gas
produced by the sublimation of water ice. This idea is based on the fact that the orbits of the
currently known MBCs are in the outer part of the asteroid belt where water ice on the
surface of asteroids can survive for billions of years when covered by a layer of dust
(\cite[Schorghofer 2008]{Schorghofer08}). A collision between an MBC and an object,
even as small as a m-sized boulder, can expose this ice. When such an MBC, with a
locally exposed sub-surface ice, approaches its perihelion, the ice sublimates and 
produces a weak atmosphere which lifts and carries dust particles from the surface 
of the MBC, giving it a cometary appearance.  

\begin{figure}
\vspace*{0.2 cm}
\begin{center}
 \includegraphics[width=4.75in]{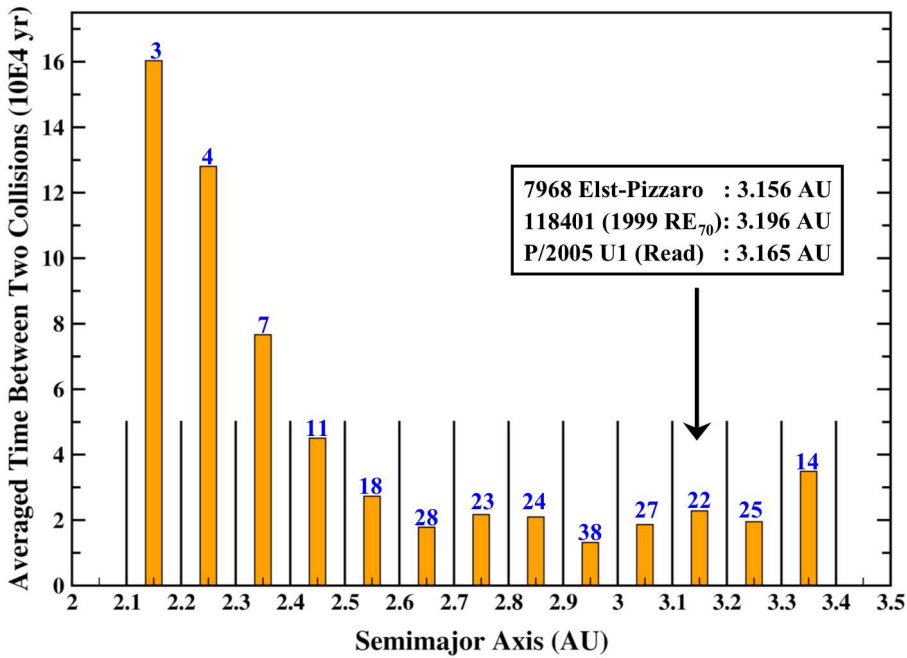} 
% \vspace*{-1.0 cm}
 \caption{Graph of the averaged time between two successive collisions of m-sized objects
 with an MBC in the orbit of 7968 Elst-Pizzaro. The numbers on top of each bar indicate the 
 percentage of the boulders of that region that collided with the MBC. As shown here, most
 of the collisions come from the vicinity of 7968 Elst-Pizzaro. The grand average of the time
 between two successive collision is approximately 40,000 years.}
  \label{fig4}
\end{center}
\end{figure}

The number of m-sized boulders and the frequency of such collisions are not exactly known.
However, it is possible to develop a simple computational model that can impose an upper
limit to these collisions. In doing so, a heuristic model was developed based on the following assumptions.

\begin{enumerate} 

\item The asteroid belt was assumed to consist of only one asteroid, 7968 Elst-Pizzaro, and
a disk of m-sized bodies. The surface density of the disk was set to have a $r^{-3/2}$ profile.

\item The accumulative size distribution $(N)$ of objects with diameter $(D)$ was considered 
to be given by $N \propto {D^n}$, where $n$ can have a value between -2 and -4. 
Following \cite[Dohnanyi (1969)]{Dohnanyi69}, it was assumed that $n=-2.5$. 

\item A total of $10^6$ m-sized boulder were randomly distributed throughout the asteroid belt.
The eccentricity of these objects were chosen from a range of 0 to 0.5, and their inclinations 
were taken to be between 0 and $25^\circ$. 

\end{enumerate}

The orbits of the m-sized objects and that of the 7968 Elst-Pizzaro were integrated for 10 Myr. Similar
to the previous simulations, integrations included all planets and Pluto. Results indicated that
on average, one m-sized object collides with this MBC every 40,000 years. As shown in Fig. 4,
a larger number of the colliding boulder come from the vicinity of 7968 Elst-Pizzaro. It is important
to emphasize that this model is simplistic, and the results represent a high upper limit. 
In a more realistic model, the numbers of large bodies and the small boulders are much higher. 
As a result, many of the m-sized objects collide with their neighboring asteroids, or are ejected 
from the asteroid belt. It is expected that in such cases, the frequency of collisions between  
km-sized MBCs and m-sized boulders to decrease to approximately one every few thousand 
years.

\section{Conclusions}

\begin{itemize}

\item Current MBCs seem to have formed through the collision and break up of bigger 
asteroids. The results of the simulations of the dynamic of these objects point to the 
Themis family as the origin of 7968 Elst-Pizzaro, 118401 (176P/LINEAR), 
and P/2005 U1 (Read).

\item Interaction with giant planet might have scattered MBCs from their original orbits
to other locations in the asteroid belt. P/2008 R1 (Garradd) seems to be one of such
scattered MBCs.

\item More MBCs may exist in low inclinations and low eccentricities in the vicinity of 
asteroid families in the outer region of the asteroid belt.

\item Collisions with  small objects might have activated MBCs or eroded them.

\item Many MBCs might have been active in the past and are either no longer active, or
will become active if hit by a small body again.
 
\item Many MBCs, with locally exposed sub-surface ice, may still be on their ways to 
their perihelion distances where they become active, or they may be awaiting collisions 
with smaller objects to get activated.

\item All sky surveys such as Pan STARRS will be able to detect more MBC in near future.

\end{itemize}

\acknowledgement
I gratefully acknowledge fruitful discussions
with H. Hsieh, D. Jewitt, H. Levison, K. Meech, D. Nesvorny, 
and N. Schorghofer. This work was partially supported by the NASA 
Astrobiology Institute under Cooperative Agreement NNA04CC08A at the 
Institute for Astronomy, NASA Astrobiology Central, the office of the 
Chancellor of the University of Hawaii, and a Theodore Dunham J. grant 
administered by Funds for Astrophysics Research, Inc. I am also grateful to
Newton's Institute for Mathematical Science at the Cambridge University
for their great hospitality during the preparation of this manuscript.

\end{document}